\documentclass[pra,twocolumn,superscriptaddress,amsmath,amssymb,floatfix]{revtex4-1}
\usepackage{graphicx}
\usepackage{eqnarray}

\begin{document}

\title{Optical conductivity of the Weyl semimetal NbP}

\author{David Neubauer}
\affiliation{1. Physikalisches Institut, Universit\"at Stuttgart,
70569 Stuttgart, Germany}
\author{Alexander Yaresko}
\affiliation{Max-Planck-Institut f\"ur Festk\"orperforschung,
Heisenbergstr. 1, 70569 Stuttgart, Germany}
\author{Weiwu Li}
\affiliation{1. Physikalisches Institut, Universit\"at Stuttgart,
70569 Stuttgart, Germany}
\author{Anja L\"ohle}
\affiliation{1. Physikalisches Institut, Universit\"at Stuttgart,
70569 Stuttgart, Germany}
\author{Ralph H\"ubner}
\affiliation{1. Physikalisches Institut, Universit\"at Stuttgart,
70569 Stuttgart, Germany} \affiliation{Institut f\"ur Funktionelle
Materie und Quantentechnologien, Universit\"at Stuttgart, 70569
Stuttgart, Germany}
\author{\\Micha B. Schilling} \affiliation {1. Physikalisches
Institut, Universit\"at Stuttgart, 70569 Stuttgart, Germany}
\author{Chandra Shekhar}
\affiliation{Max-Planck-Institut f\"ur Chemische Physik fester
Stoffe, 01187 Dresden, Germany}
\author{Claudia Felser}
\affiliation{Max-Planck-Institut f\"ur Chemische Physik fester
Stoffe, 01187 Dresden, Germany}
\author{Martin Dressel}
\affiliation{1. Physikalisches Institut, Universit\"at Stuttgart,
70569 Stuttgart, Germany}
\author{Artem V. Pronin}
\affiliation{1. Physikalisches Institut, Universit\"at Stuttgart,
70569 Stuttgart, Germany}

\date{March 26, 2018}

\begin{abstract}
The optical properties of (001)-oriented NbP single crystals have
been studied in a wide spectral range from 6~meV to 3~eV from room
temperature down to 10~K. The itinerant carriers lead to a
Drude-like contribution to the optical response; we can further
identify two pronounced phonon modes and interband transitions
starting already at rather low frequencies. By comparing our
experimental findings to the calculated interband optical
conductivity, we can assign the features observed in the measured
conductivity to certain interband transitions. In particular, we
find that transitions between the electronic bands spilt by
spin-orbit coupling dominate the interband conductivity of NbP below
100~meV. At low temperatures, the momentum-relaxing scattering rate
of the itinerant carriers in NbP is very small, leading to
macroscopic characteristic length scales of the momentum relaxation
of approximately 0.5~$\mu$m.
\end{abstract}

\maketitle

\section{Introduction}

NbP is a nonmagnetic non-centrosymmetric Weyl semimetal (WSM) with
extremely large magnetoresistance and ultrahigh carrier mobility
\cite{Shekhar2015, Wang2016}. These extraordinary transport
properties are believed to be caused by quasiparticles in chiral
Weyl bands. According to band-structure calculations NbP possesses
24 Weyl nodes, i.e.\ twelve pairs of the nodes with opposite
chiralities \cite{Huang2014, Weng2015, Sun2015, Lee2015}. The nodes
are ``leftovers'' of nodal rings, which are gapped by spin-orbit
coupling (SOC) everywhere in the Brillouin zone (BZ), except of
these special points \cite{Huang2014, Weng2015, Lee2015, Ahn2015}.
The nodes can be divided in two groups, commonly dubbed as W1 (8
nodes) and W2 (16 nodes). Most recent band-structure calculations
agree well on the energy position of the W1 nodes: 56 - 57~meV below
the Fermi level $E_{F}$~\cite{Klotz2016, Wu2017, Grassano2018}; the
position of the W2 nodes is specified less accurately, ranging from
5~\cite{Klotz2016, Wu2017} to 26 meV~\cite{Grassano2018} above
$E_F$. Furthermore, the W2 cones could be strongly tilted along a
low-symmetric direction in BZ~\cite{Wu2017}, realizing thus a
type-II WSM state~\cite{Soluyanov2015}.

The free-carrier dynamics of NbP has attracted considerable
attention due to the possibility of hydrodynamic election behavior.
Recently, a so-called axial-gravitational anomaly, relevant in the
hydrodynamic regime~\cite{Lucas2016}, was reported in
NbP~\cite{Gooth2017NbP}. In turn, studying the hydrodynamic behavior
of electrons is very interesting because the conduction of viscous
electron fluids can be extremely high, exceeding the fundamental
ballistic limit \cite{Gurzhi1963,Gurzhi1968,Guo2017}. Dirac
materials with highly mobile electrons are prime candidates for
realizing this super-ballistic conductivity. It was recently
reported that electrons flow in a hydrodynamic fashion in clean
samples of graphene \cite{Bandurin2016,Crossno2016} and WP$_{2}$
\cite{Gooth2017WP2}; and the higher-than-ballistic conduction was
found through graphene constrictions \cite{Kumar2017}.

Similar to all WSM, the physical properties of NbP are determined by
their low-energy electron dynamics, including both, free-carrier
response and interband transitions \cite{Wehling2014}. Infrared
optical methods enable direct access to this dynamics. For example,
the interband optical response of a single isotropic
three-dimensional Weyl band, being expressed in terms of the real
part of the complex conductivity, should follow a linear frequency
dependence with the pre-factor given by the band Fermi velocity
$v_F$~\cite{Hosur2012, Bacsi2013, Ashby2014}:
\begin{equation}
\sigma(\omega) = \frac{e^2} {12 h} \frac{\omega} {v_F} \quad .
\label{simple}
\end{equation}
Here, electron-hole symmetry is assumed, and the complex conductivity
is $\hat{\sigma}(\omega) = \sigma(\omega) - {\rm
i}\omega(\varepsilon(\omega)-1)/4\pi$, with $\varepsilon(\omega)$ the
real part of the dielectric function. For $N_\text{W}$ identical
Weyl bands, the right side of Eq.~(\ref{simple}) should be
multiplied by $N_\text{W}$. Such a linear behavior of
the optical conductivity
($\sigma\propto\omega$) has indeed been observed in a number of
well-established and proposed three-dimensional   Weyl/Dirac-semimetal
systems~\cite{Timusk2013, Chen2015, Sushkov2015, Xu2016,
Neubauer2016, Ueda2016, Kimura2017, Huett2018}.

In the hydrodynamic regime, the rates for momentum-relaxing
scattering $\Gamma_{\textrm{mr}} = 1/\tau_{\textrm{mr}}$ and
momentum-conserving scattering $\Gamma_{\textrm{mc}} =
1/\tau_{\textrm{mc}}$ of the itinerant carriers differ appreciably:
$\Gamma_{\textrm{mc}} \gg \Gamma_{\textrm{mr}}$ \cite{Gurzhi1963,
Gurzhi1968, Scaffidi2017, Lucas2017}. The momentum-relaxing
scattering manifests itself in the optical-conductivity
spectra~\cite{Dressel2002}, thus one can directly determine the
corresponding scattering time $\tau_{\textrm{mr}}$ from
$\sigma(\omega)$.

For this paper, we have measured and analyzed both, interband and
itinerant-carrier, conductivity of NbP. We show that the low-energy
interband conductivity of NbP is dominated by transitions between
the bands with parallel dispersions that are split by spin-orbit
coupling. These excitations, as well as the Drude response of the
itinerant carriers, completely mask the linear-in-frequency
$\sigma(\omega)$ due to the three-dimensional chiral Weyl bands. At
somewhat higher frequencies (1400 - 2000~cm$^{-1}$, 175 - 250~meV),
$\sigma(\omega)$ becomes roughly linear. Our calculations
demonstrate that this linearity stems from the fact that all
electronic bands, which are involved in the transitions with
relevant energies, are roughly linear. In addition, we find that at
low temperatures the itinerant carriers in at least one of the
conduction channels possess an extremely long momentum-relaxing
scattering time; the characteristic length scale of momentum
relaxation, $\ell_{\textrm{mr}} = v_F\tau_{\textrm{mr}}$, is then
basically macroscopic, supporting that the hydrodynamic regime can
be realized in NbP.

\section{Samples Preparation, Experimental and Computational Details}

Single crystals of NbP were synthesized according to the description
reported in Refs.~\cite{Martin1990, Shekhar2015}: a polycrystalline
NbP powder was synthesized in a direct reaction of pure niobium and
red phosphorus; the single NbP crystals were grown from the powder
via vapor-transport reaction with iodine.

\begin{figure}[b]
\centering
\includegraphics[width=0.9\columnwidth]{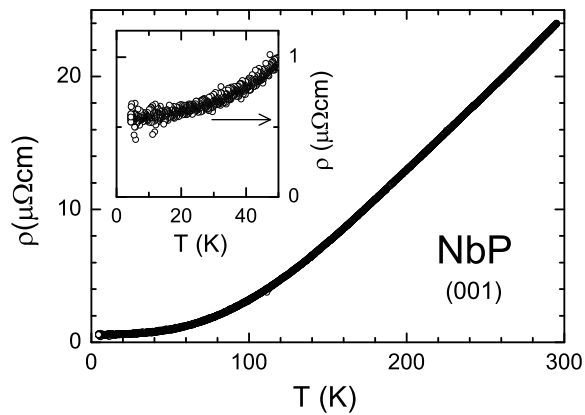}
\caption{Temperature-dependent (001)-plane dc resistivity of the NbP
sample used in the optical measurements.} \label{fig:res}
\end{figure}

The electrical resistivity, $\rho_{dc} (T)$, was measured as a
function of temperature within the (001)-plane in four-contact
geometry. The experiments were performed on a small piece, cut from
the specimen used for the optical investigations. The results of the
dc measurements are plotted in Fig.~\ref{fig:res}. A clear metallic
behavior with linear-in-temperature resistivity is observed down to
approximately 100~K. Below, $\rho_{dc} (T)$ levels off and
approaches the residual resistivity $\rho_0$ = $0.55~\mu\Omega$cm;
the resistivity ratio is $\rho$($300$ K)/$\rho_0 = 40$. These values
are well comparable with the ones reported in
literature~\cite{Shekhar2015, Wang2016, Zhang2015, Wang2015c}. Note,
NbP exhibits an extremely high mobility independent on the residual
resistivity ratio~\cite{Wang2015c}.

The normal-incidence optical reflectivity $R(\nu,T)$ was measured on
the (001)-surfaces of a large ($2~{\rm mm} \times 2~{\rm mm}$ in
lateral dimensions) single crystal from room temperature down to
$T=10$~K covering a wide frequency range from $\nu = \omega/(2\pi
c)= 50$ to 12\,000~cm$^{-1}$. The temperature-dependent experiments
were supplemented by room-temperature reflectivity measurements up
to 25\,000~cm$^{-1}$. In the far-infrared spectral range below
700~cm$^{-1}$, a Bruker IFS 113v Fourier-transform spectrometer was
employed with \textit{in situ} gold coating of the sample surface
for reference measurements. At higher frequencies, we used a Bruker
Hyperion infrared microscope attached to a Bruker Vertex 80v
spectrometer. Here, freshly evaporated gold mirrors (below
12\,000~cm$^{-1}$) and protected silver (above 12\,000~cm$^{-1}$)
served as reference.

For the Kramers-Kronig analysis~\cite{Dressel2002} we involved the
x-ray atomic scattering functions for high-frequency
extrapolations~\cite{Tanner2015}. From recent optical investigations
of materials with highly mobile carriers, it is known
\cite{Schilling2017Yb, Schilling2017Zr} that the commonly applied
Hagen-Rubens extrapolation to zero frequency is not adequate: the
very narrow zero-frequency component present in the spectra
corresponds to a scattering rate comparable to (or even below) our
lowest measurement frequency, $\nu_{\rm min} \approx 50$~cm$^{-1}$.
Thus, we first fitted the spectra with a set of Lorentzians (similar
fitting procedures can be utilized as a substitute of the
Kramers-Kronig analysis~\cite{Kuzmenko2005, Chanda2014}) and then we
used the results of these fits between $\nu = 0$ and $\nu_{\rm min}$
as zero-frequency extrapolations for subsequent Kramers-Kronig
transformations. We note that our optical measurements probe the
bulk material properties, as the penetration depth exceeds 20~nm for
any measurement frequency.

We performed band structure calculations within the local density
approximation (LDA) based on the crystal structure of NbP determined
by experiments \cite{XGEH+96}; we employed the linear muffin-tin
orbital method \cite{And75} as implemented in the relativistic PY
LMTO computer code. Some details of the implementation can be found
in Ref.\ \onlinecite{book:AHY04}. The Perdew-Wang parametrization
\cite{PW92} was used for the exchange-correlation potential. SOC was
added to the LMTO Hamiltonian in the variational step. BZ
integrations were done using the improved tetrahedron method
\cite{BJA94}. Dipole matrix elements for interband optical
transitions were calculated on a $96 \times 96 \times 96$ $k$-mesh
using LMTO wave functions. As was shown in Ref.\
\onlinecite{Chaudhuri2017}, it is necessary to use sufficiently
dense meshes in order to resolve transitions between the SOC-split
bands. The real part of the optical conductivity was calculated by
the tetrahedron method.

\section{Results and Analysis}

\begin{figure}[t]
\centering
\includegraphics[width=0.8\columnwidth]{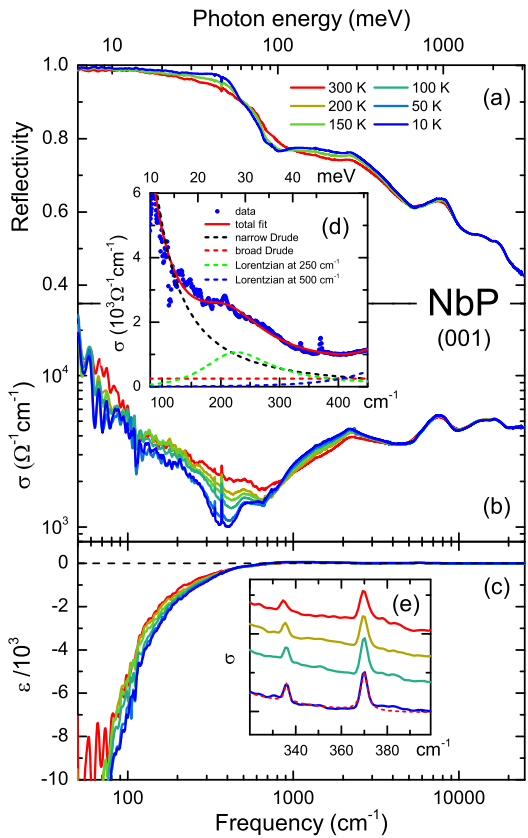}
\caption{(a) Optical reflectivity, real parts of the (b)  optical
conductivity and (c) dielectric permittivity of NbP at selected
temperatures between $T=10$ and 300~K; note the logarithmic
frequency scale. The inset (d) shows a simple fit of the low-energy
$\sigma(\nu)$ at $T=10$~K by the sum of two Drude terms (narrow and
broad) and two Lorentzians centered at 250 and 500~cm$^{-1}$, which
mimic the interband transitions. The inset (e) displays the phonon
modes in $\sigma(\nu)$ on an enlarged frequency scale. The dashed
red line corresponds to a fit of the features at $T=10$~K by two
narrow Lorentzians centered at 336 and 370~cm$^{-1}$, respectively.}
\label{ref_sig_eps}
\end{figure}

Fig.~\ref{ref_sig_eps} displays the overall reflectivity $R(\nu)$,
the real part of the conductivity $\sigma(\nu)$, and the dielectric
constant $\varepsilon(\nu)$ of NbP for different temperatures. For
frequencies higher than 5000~cm$^{-1}$, the optical properties are
basically independent on temperature. In the spectra we can identify
the signatures of: (i)~phonons, (ii)~itinerant-carrier (intraband)
absorption, and (iii)~interband transitions. Below we discus these
spectral features separately.

in TaAs, for example, such a Fano resonance reportedly signals a
strong coupling between phonons and electronic transitions [50]).

\subsection{Phonons}

In the far-infrared range, two sharp phonon peaks can be seen in
Fig.~\ref{ref_sig_eps} at 336 and 370~cm$^{-1}$. Due to their
symmetric shape, they can be nicely fitted with Lorentzians, as
demonstrated in panel (e). This is in contrast to the Fano-shape
phonon resonances observed in TaAs, where the asymmetric line shape
reportedly signals a strong coupling between phonons and electronic
transitions~\cite{FanoTaAs}. In NbP, four infrared-active phonons
are expected \cite{Chang2016}; however, there is no full consensus
on the calculated frequencies \cite{Chang2016, Liu2016}. The phonon
positions observed in our spectra agree very well with the
calculations from Ref.~\cite{Liu2016} as well as with the Raman data
presented there (in non-centrosymmetric structures, same phonon
modes can be both, infrared and Raman, active). Thus, following
Ref.~\cite{Liu2016}, we assign the observed features at 336 and
370~cm$^{-1}$ to those lattice vibrations that mainly involve the
light P atoms. The other two infrared-active phonon modes are
apparently too weak to be resolved on the electronic background.

\subsection{Itinerant charge carriers}

At the lowest frequencies, NbP exhibits an optical response typical
for metals, i.e. the itinerant carriers dominate: the reflectivity
approaches unity, $\varepsilon(\omega)$ is negative and diverges as
$\omega \rightarrow 0$, $\sigma(\omega)$ exhibits a narrow
zero-frequency peak, as seen in Fig.~\ref{ref_sig_eps}~(a-c). Panel
(d) clearly shows a shoulder on this peak that can be fitted by a
Lorentzian; it will be discussed in the next section.

For fitting the optical spectra at our lowest frequencies, we use
the measured dc-conductivity values as the zero-frequency limit of
the Drude term~\cite{Dressel2002}, and we allow both, the plasma
frequency and the scattering rate, to vary freely. The best fit to
the data is then obtained with a momentum-relaxing scattering rate,
$\gamma_{\textrm{mr}} = 1/(2\pi c \tau_{\textrm{mr}})$, as low as
4.5 cm$^{-1}$ at $T=10$~K. Thus, $\tau_{\textrm{mr}} (10 \textrm{K})
= 1.2$ ps and the momentum-relaxation length, $\ell_{\textrm{mr}} =
v_\textrm{F}\tau_\textrm{mr}$, becomes as long as 0.2 to 0.6~$\mu$m.
Here we utilized the lower, $1.5\times 10^{5}$ m/s, and,
respectively, the upper, $4.8 \times 10^{5}$ m/s, boundaries for the
(001)-plane Fermi velocity $v_\textrm{F}$ obtained from
experiment~\cite{Shekhar2015, Wang2016} and theory~\cite{Lee2015,
Ahn2015}. At elevated temperatures, $\gamma_{\textrm{mr}}$ rises,
reaching 35~cm$^{-1}$ at $T=300$~K. This corresponds to
$\ell_{\textrm{mr}}$ of 20 to 70~nm.

Since the typical sample size for ballistic transport measurements
is around 0.1 to 1~$\mu$m~\cite{Kumar2017, Datta1995}, one can
actually realize ballistic conduction in NbP at low temperatures.
For the hydrodynamic electron behavior discussed in the
introduction, the energy-relaxing momentum-conserving
electron-electron scattering must happen on time scales shorter than
$\tau_{\rm{mr}}$ \cite{Gurzhi1963, Gurzhi1968, Scaffidi2017,
Lucas2017}. Because the found $\tau_{\textrm{mr}}$ is rather large,
the hydrodynamic behavior and, eventually, the super-ballistic
electron flows may be possible in NbP.

Less mobile carriers from parabolic bands~\cite{Shekhar2015,
Sun2015, Ahn2015, Klotz2016, Wu2017}, however, can obscure the
observation of this behavior. Thus, a proper adjustment of the Fermi
level is crucial, as can be achieved, e.g., by Gd doping
\cite{Niemann2017}. Fingerprints of these low-mobility carriers can
be identified in the optical conductivity of undoped NbP as another,
very broad and relatively weak, Drude band. Such two-channel optical
conductivity has been recently observed, for instance, in
YbPtBi~\cite{Schilling2017Yb}. In the case of NbP, this broad Drude
term is present in the far-infrared spectral range as a basically
frequency-independent pedestal in $\sigma(\omega)$, clearly seen in
Fig.~\ref{ref_sig_eps}(d).

\begin{figure}[t]
%\centering
\includegraphics[width=\columnwidth]{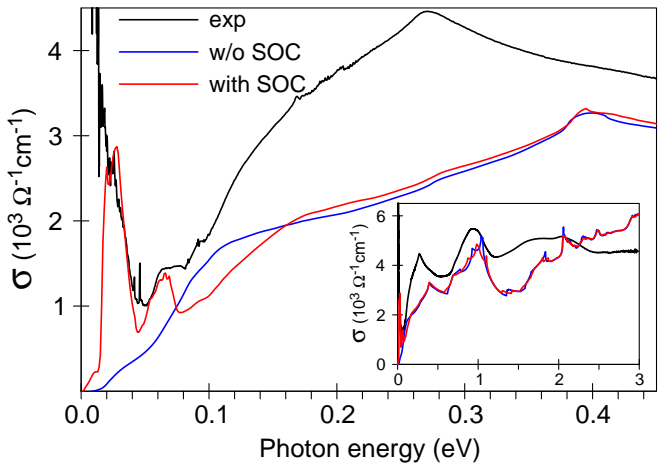}
\caption{\label{fig:sigma}Interband optical conductivity of NbP
calculated with (red line) and without (blue line) SOC and the total
experimental NbP conductivity at 10 K (black line). Intraband
(Drude) contributions to the conductivity are not included in the
computations. Inset shows same sets of data on a broader photon
energy scale (0 -- 3 eV). All spectra are for the (001)-plane
response.}
\end{figure}

\subsection{Interband transitions: optical experiments and calculations based on the electronic band
structure}

Calculations of the interband optical conductivity based on
calculated electronic band structure are very useful, but seem to be
rather challenging in (topological) semimetals. A survey of the
available literature reveals only a qualitative match between the
calculated optical conductivity and experimental results
\cite{Grassano2018, Kimura2017, Frenzel2017, Chaudhuri2017}. In the
most interesting low-energy part of the spectrum (less than a few
hundred meV), a reasonable agreement is particularly hard to achieve
\cite{Kimura2017, Chaudhuri2017}. In the case of NbP, we reach a
fairly good qualitative match between our calculations of the
optical conductivity and experimental spectra even at low energies;
this allows us to identify the origin of various spectral features
observed in  $\sigma(\omega)$.

In Fig.\ \ref{fig:sigma} we compare the experimental low-temperature
optical conductivity to the interband (001)-plane conductivity
calculated with and without SOC. Note, that the itinerant-carriers
contributions have not been subtracted from the experimental
conductivity. The calculated spectra qualitatively agree with the
experiment: the inset in Fig.\ \ref{fig:sigma} illustrates that the
peaks and deeps in the calculated $\sigma(\omega)$ reasonably
coincide with those in the experimental data. The calculated spectra
contain more fine structures than the experimental $\sigma(\omega)$,
as no broadening was applied to the computed spectra in order to
simulate finite life-time effects. The experimental $\sigma(\omega)$
in general falls above the calculated interband conductivity.
Partly, this can be due to the broad Drude contribution present in
the experimental spectra.

Above an energy of 0.2~eV, the effect of SOC on the theoretical
$\sigma(\omega)$ is negligible; but for smaller energies the spectra
computed with and without taking SOC into account differ
significantly. The interband contribution to the conductivity
calculated without SOC increases smoothly when raising the photon
energy to 0.1~eV. When SOC is included, however, two sharp peaks
appear around 30 and 65~meV (corresponding to $\sim 250$ and
500~cm$^{-1}$). The latter matches very well the shoulder we
observed on the narrow Drude term, as shown in
Fig.~\ref{ref_sig_eps}(d). The former feature can be directly
associated with the bump observed at all temperatures at around
500~cm$^{-1}$ in the  measured spectra plotted in
Fig.~\ref{ref_sig_eps}(b). The fact that these two peaks appear only
in those calculations including SOC indicates that they must be
related to the transitions between the SOC-split bands.

\begin{figure*}[t]
\centering
\includegraphics[width=\textwidth]{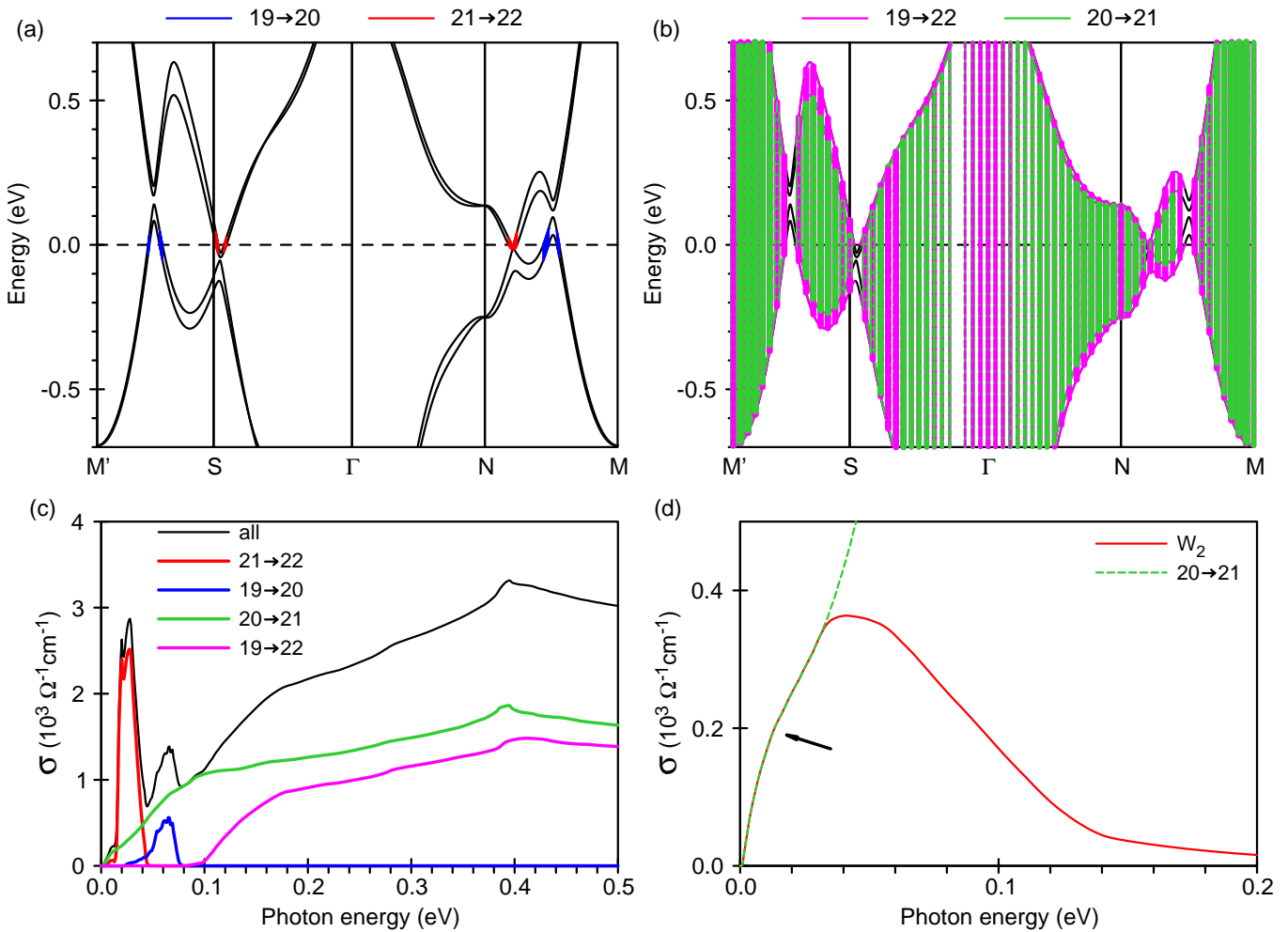}
\caption{\label{fig:transitions} (a,b)~Band structure of NbP along
selected lines in the BZ with allowed transitions shown as colored
vertical lines. The thickness of the lines is proportional to the
transition probability. (c)~Calculated contributions to the
interband conductivity from transitions between different pairs of
bands crossing $E_F$, as indicated, and the total calculated
interband conductivity of NbP. (d)~Contribution to $\sigma_{20\to
21}(\omega)$ from the transitions within the small volumes in
$\mathbf{k}$-space, enclosing the W2 Weyl points (red solid line,
see text for details), in comparison with the total $\sigma_{20\to
21}(\omega)$ (dashed green line). The arrow indicates a kink, which
corresponds to the point, where the chiral Weyl bands merge.}
\end{figure*}

\begin{figure}[b]
\centering
\includegraphics[width=0.9\columnwidth]{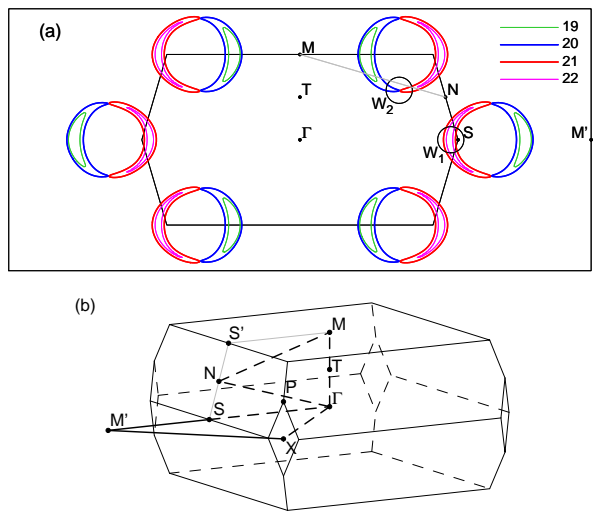}
\caption{\label{fig:fs}(a) Fermi surface cross sections by $k_y=0$
plane. Small black circles illustrate integration volumes around
Weyl points. (b) BZ of NbP. Weyl points from the W1 set are situated
near the S point, while the points from the W2 set are close to the
N--M line.}
\end{figure}

Our conclusion gets support when decomposing the calculated
$\sigma(\omega)$ into contributions coming from transitions between
different pairs of bands crossing $E_{\rm F}$. When SOC is
neglected, two doubly degenerate bands with predominant Nb $d$
character cross on a $k_{x/y}=0$ mirror plane and form one electron
and one hole Fermi surface with crescent-shaped cross sections by
this plane \cite{Lee2015}. This degeneracy is lifted when SOC is
accounted for; thus, four non-degenerate bands, numbered 19 to 22,
now cross $E_{\rm F}$ as shown in Figs.\ \ref{fig:transitions} and
\ref{fig:fs} (at every given \textbf{k} point, the bands are
numbered with increasing energy). The band structure of NbP,
calculated along selected lines in the BZ [cf.\ Fig.\
\ref{fig:fs}(b)], and the allowed transitions between different
bands are shown in Figs.\ \ref{fig:transitions}(a,b). Here, the
thickness of the vertical lines connecting occupied initial and
unoccupied final states is proportional to the probability of the
interband transition at a given $\mathbf{k}$ point. In panel (c) of
Fig.\ \ref{fig:transitions}, we plot the various contributions to
the total interband conductivity from the individual interband
transitions $19 \to 20$, $21 \to 22$, $19 \to 22$, and $20 \to 21$.
The contributions of the two remaining transitions, $19 \to 21$, $20
\to 22$, are much smaller. Thus, the total low-energy conductivity
$\sigma(\omega)$ is basically the sum of the four contributions
shown in Fig.\ \ref{fig:transitions}(c).

The pronounced narrow peak at 30~meV is solely formed by transitions
between the bands 21 and 22, while the 65~meV mode is due to the $19
\to 20$ transitions; i.e.\ both features stem from transitions
between SOC-split bands. These transitions are only allowed in a
small volume of the $\mathbf{k}$-space, where one of the two
SOC-split bands is occupied, while the other one is empty (see the
red and blue lines in Fig.\ \ref{fig:transitions}(a)), i.e., between
the nested crescents in Fig.\ \ref{fig:fs}(a). Within this volume,
the SOC-split bands are almost parallel to each other. This ensures
the appearance of strong and narrow peaks in the joint density of
states for the corresponding interband transitions. The peak
positions are determined by the average band splitting, which, in
turn, is of the order of the SOC strength of the Nb $d$ states,
$\xi_{d} \approx 85$~meV. Finally, since the dipole matrix elements
for these transitions are rather large, the two peaks dominate the
low-energy interband conductivity.

Besides these two peaks, there are also contributions to the optical
conductivity with a smooth $\omega$ dependence. These contributions
originate in transitions between the touching bands 20 and 21 [green
lines in Fig.\ \ref{fig:transitions}(b,c)] and between the bands 19
and 22, which are separated by a finite gap everywhere in the BZ
[magenta lines]. Accordingly, $\sigma_{20\to21}(\omega)$ starts at
zero energy, while $\sigma_{19\to22}(\omega)$ at 0.1~eV. Both
conductivity contributions, $\sigma_{20\to21}$ and
$\sigma_{19\to22}$, increase, when the photon energy rises from 0 to
0.4 eV. Both contributions, $\sigma_{19\to22}$ and
$\sigma_{19\to22}$, as well as the total calculated $\sigma(\omega)$
exhibit sharp kinks (Van Hove singularities~\cite{Yu2010}) at
0.4~eV, which are related to the transitions between flat parallel
bands near the N point, see Fig.\ \ref{fig:transitions}(b). The
experimental $\sigma(\omega)$ demonstrates such a kink at somewhat
lower energy, 0.27 eV (Fig.\ \ref{fig:sigma}); still, we find this
match reasonable.

In the vicinity of a Weyl point, the optical conductivity is
expected to be proportional to frequency, Eq.~(\ref{simple}). The
two sets of Weyl points in NbP, W1 and W2, are formed by touching
points of the bands 20 and 21. In agreement with previous
results~\cite{Lee2015, Klotz2016, Wu2017, Grassano2018}, our LMTO
calculations yield the W1 points approximately 50~meV below $E_F$.
Consequently, their contribution to the conductivity cannot start at
zero frequency. On the other hand, the energy of the W2 points in
the present calculations is very close to $E_F$ and, thus,
transitions near W2 may provide a linearly vanishing
$\sigma(\omega)$ as $\omega \rightarrow 0$. To verify this behavior,
we calculate the contribution to $\sigma_{20\to 21}(\omega)$ from a
$\mathbf{k}$ volume with a radius of $\sim 0.05 \frac{2\pi}{a}$ ($a$
is the in-plane lattice constant) around the averaged position of a
pair of W2 points. The contribution indeed shows a linear $\omega$
dependence as $\omega \rightarrow 0$, see Fig.\
\ref{fig:transitions}(d). The smooth kink at $\sim 15$~meV, marked
with an arrow, corresponds to the merging point of the chiral Weyl
bands~\cite{Yu2010, Tabert2016}. We should note that in experiments,
this linear interband optical conductivity at low $\omega$ is
completely masked by the itinerant carriers and by the strong peaks
due to the transitions between the SOC-split bands, as discussed
above.

Although no linear-in-frequency $\sigma(\omega)$ due to the
transitions within the chiral Weyl bands can be seen in NbP at low
frequencies, both, experimental and computed, $\sigma(\omega)$
demonstrate a sort of linear increase with $\omega$ at higher
frequencies: 180 to 250~meV for the experimental and up to 360~meV
for the calculated optical conductivity, see Fig.\ \ref{fig:sigma}.
As apparent from our calculations, the linearity just reflects the
fact that all the electronic bands, involved in the transitions with
corresponding energies, are roughly linear (but not parallel to each
other), see Fig.\ \ref{fig:transitions}(b).

\begin{figure}[t]
\centering
\includegraphics[width=0.9\columnwidth]{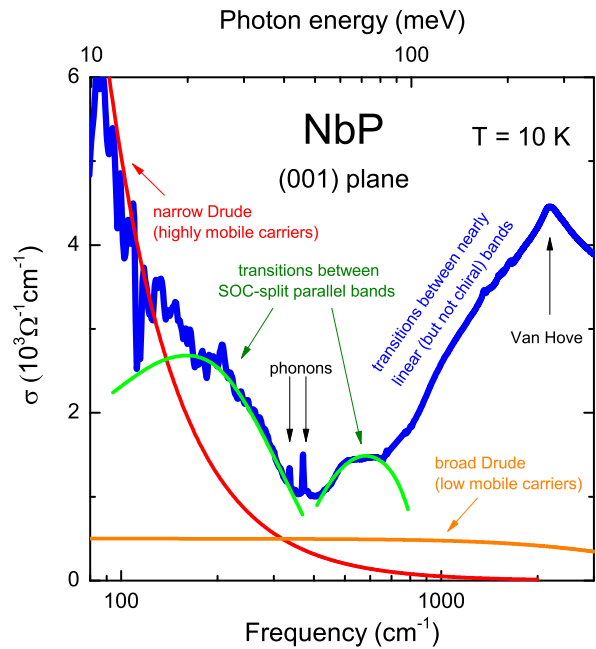}
\caption{Low-frequency portion of the real part of
NbP optical conductivity at 10 K and assignment of the observed
features to different absorption mechanisms, as discussed in the
course of this paper.} \label{sigma010K}
\end{figure}

Based on the comparison between the calculated and the experimental
conductivity, we can assign the observed spectral features to
different absorption mechanisms. Fig.~\ref{sigma010K} schematically
summarizes these assignments.

Before we conclude, we would like to emphasize the importance of the
transitions between the SOC-split bands. So far, the strong
influence of these transitions on the low-energy conductivity of
WSMs has not been fully appreciated. Using a modified Dirac
Hamiltonian~\cite{Burkov2011}, Tabert and Carbotte \cite{Tabert2016}
calculated the optical conductivity for a four-band model, relevant
for many WSMs. In this model, the band structure consists of four
isotropic non-degenerate three-dimensional bands, two of which cross
and the other two are gapped. The band structure is mirror symmetric
in energy with respect to the Weyl nodes and the Weyl cones are
anisotropic in $\mathbf{k}$-space at low energies. This model
definitely grasps the main features of the band structure of many
WSMs, including those from the TaAs family; but -- apart from
neglecting possible band anisotropy -- it does not take into account
the transitions between the SOC-split bands: these transitions are
considered forbidden in the model, while in the real WSMs they might
play an important role, as we have shown for NbP.

Finally, we would like to note that TaAs and TaP demonstrate sharp
absorption peaks at frequencies, which compare well to those of the
transitions between the SOC-split bands in NbP, cf.\ Fig. 3 from
Ref.~\cite{Kimura2017}. Previously, these features have been
assigned to transitions between the merging (saddle) points of the
Weyl bands, even though, e.g., in TaP this assignment is at odds
with the absence of chiral carriers~\cite{Arnold2016}. We suggest to
reconsider this assignment.

\section{Conclusions}

We measure and analyze the interband and itinerant-carrier optical
conductivity of NbP. From the electronic band structure, we
calculate the interband optical conductivity and decompose it into
contributions from the transitions between different bands. By
comparing these contributions to the spectral features in the
experimental conductivity, we assign the observed features to
certain interband transitions. We argue that the low-energy (below
100~meV) interband conductivity is dominated by transitions between
almost parallel bands, split by spin orbit coupling. Hence, these
transitions manifest themselves as relatively sharp peaks centered
at 30 and 65~meV. These peaks and the low-energy itinerant-carrier
conductivity (Drude-like) conceal the linear-in-frequency
contribution to $\sigma(\omega)$ from the transitions within the
chiral Weyl bands. Our calculations demonstrate that the nearly
linear in $\omega$ conductivity at around 200 -- 300~meV is
naturally explained by the fact that all electronic bands, involved
in the transitions with such frequencies, possess approximately
linear dispersion relations. We also identify two optical phonons
and assign them to the vibrations, which mostly involve P atoms.
Finally, we find that the carriers in one of the conduction channels
possess extremely low momentum-relaxing scattering rates at low
temperatures, leading to macroscopic characteristic length scales of
momentum relaxation of about 0.5~$\mu$m at $T=10$~K.

\section{Acknowledgements}

We thank G. Untereiner and S. Prill-Diemer for experimental support,
and U. S. Pracht and H. B. Zhang for fruitful discussions. This
project was funded by the Deutsche Forschungsgesellschaft (DFG) via
grant No. DR228/51-1.

\end{document}